\documentclass{jpsj-suppl}
\usepackage{txfonts} 
\usepackage{wrapfig}

\title{Strong Orientation Dependence of Multinucleon Transfer Processes in $^{238}$U+$^{124}$Sn Reaction}

\author{Kazuyuki \textsc{Sekizawa}$^{1}$ and Kazuhiro \textsc{Yabana}$^{1,2}$}

\inst{$^{1}$Graduate School of Pure and Applied Sciences, University of Tsukuba, Tsukuba 305-8571, Japan \\
$^{2}$Center for Computational Sciences, University of Tsukuba, Tsukuba 305-8577, Japan}

\email{sekizawa@nucl.ph.tsukuba.ac.jp}

\recdate{September 30, 2014}

\abst{
We theoretically investigate multinucleon transfer (MNT) processes
in $^{238}$U+$^{124}$Sn reaction at $E_\mathrm{lab}=5.7$
MeV/$A$ using the time-dependent Hartree-Fock (TDHF)
theory. For this reaction, measurements of MNT processes
have been reported, showing substantial MNT cross sections
accompanying more than ten protons. From the calculation,
we find that the amount of transferred nucleons depends much on
the relative orientation between the deformation axis of $^{238}$U
and the relative vector connecting centers of $^{238}$U and
$^{124}$Sn nuclei. We find a formation of thick neck when
the $^{238}$U collides from its tip with $^{124}$Sn.
However, the neck formation is substantially suppressed when
$^{238}$U collides from its side. We have found that a large
number of protons are transferred in the tip collision. This is
caused by the breaking of the neck and subsequent absorption
of nucleons in the neck region. We thus conclude that the measured
MNT processes involving about ten protons originate from the neck
breaking dynamics in the tip collisions of a deformed $^{238}$U nucleus.
}

\kword{multinucleon transfer, low-energy heavy ion reactions, TDHF}

\def\zI{\mathrm{i}\hspace{0.2mm}}

\begin{document}
\maketitle

\section{Introduction}

About 30 years ago, pioneering measurements of multi-nucleon
transfer (MNT) cross sections in $^{238}$U-induced dissipative
collisions were achieved \cite{Mayer238U}. In the measurements
of $^{238}$U + $^{124}$Sn collisions, MNT processes
accompanying more than ten protons from $^{238}$U have been
reported. Possible structural effects have been advocated to explain
the fact that lighter fragments with neutron number approximately
equal to $N=82$ are produced abundantly. Although there have
been extensive efforts to clarify the reaction mechanism both
experimentally and theoretically, the origin of the transfer of
many protons has not yet been clear.

Recently, we have developed a theoretical method to investigate
MNT processes in heavy ion reactions within the microscopic 
framework of the time-dependent Hartree-Fock (TDHF) theory. 
To calculate transfer probabilities from the TDHF wave function 
after collision, we have used a particle number projection (PNP) 
technique \cite{Projection}. From the thorough analyses reported 
in Ref.~\cite{KS_KY_MNT}, we have shown that the TDHF theory
is capable of describing MNT processes quantitatively, in comparable
accuracy to other existing theories. Since the TDHF calculation is
microscopic treating nucleons' degrees of freedom on equal footing
and does not include any empirical parameter specific to colliding
systems, we consider that it provides reliable microscopic pictures
for the reaction dynamics.

We are undertaking investigations for the $^{238}$U+$^{124}$Sn
in the TDHF theory to clarify the mechanism of MNT processes.
In this short article, we will show tentative results of our calculation.
We show that the MNT involving more than ten protons can be
described by the TDHF calculation and that a deformed structure of
$^{238}$U is crucial in the reaction dynamics.

\section{Method and Results}

To describe $^{238}$U+$^{124}$Sn collisions, we use the
computational code of TDHF calculations for nuclear collisions
which we have developed recently~\cite{KS_KY_MNT}.
We employ a uniform spatial grid in the three-dimensional
Cartesian coordinates to represent single-particle wave functions.
The grid spacing is taken to be 0.8~fm. For first and second
derivatives, we use 11-point finite difference formula. The
projectile and target nuclei are calculated using a box with
$30 \times 30 \times 30$ grid points. For reaction calculations,
we use a box with $70 \times 70 \times 30$ grid points for
non-central collisions and $90 \times 40 \times 30$ grid points
for central collisions. We choose the incident direction parallel to
the $x$-axis, and the direction of the impact parameter vector
parallel to positive-$y$ direction. The reaction plane is thus
$xy$-plane. As the initial condition, we place wave functions of
two nuclei separated by 24~fm in the incident direction. Because
the total number of protons included in projectile and target nuclei
is very large, $Z=92+50=142$, no fusion reactions has been
observed at any impact parameters. At the final stage of calculations,
there always appear two fragments, a projectile-like fragment (PLF)
and a target-like fragment (TLF). We continued time evolution
calculations until the relative distance between the two fragments
becomes larger than 28~fm. For all calculations reported in this
article, we use Skyrme SLy5 parameter set \cite{Chabanat}, 
as in Ref.~\cite{KS_KY_MNT}.

To calculate transfer probabilities, we define two spatial regions,
$V$ and $\bar{V}$. The region $V$ is taken to be a sphere around
the center-of-mass of a fragment with a radius of 14~fm. The region
$\bar{V}$ is taken to be the rest of the space. Using the PNP technique,
the probability to find $N$ nucleons with isospin $q$ ($=n$ or $p$)
in the spatial region $V$ is given by \cite{KS_KY_MNT,Projection}
\begin{equation}
P_N^{(q)} = \frac{1}{2\pi} \int_0^{2\pi}
\hspace{-1.5mm} d\theta \; e^{\zI N\theta}\det\Bigl\{ 
\bigl<\psi_i^{(q)}\big|\psi_j^{(q)}\bigr>_{\bar{V}} + e^{-\zI\theta}
\bigl<\psi_i^{(q)}\big|\psi_j^{(q)}\bigr>_V \Bigr\}.
\label{Pn_projection}
\end{equation}
Since the TDHF wave function is a direct product of two Slater
determinants for protons and neutrons, the probability to find $N$
neutrons and $Z$ protons in the spatial region $V$ is given by
a product of probabilities for neutrons and protons, $P_{N,Z} =
P_{N}^{(n)}P_{Z}^{(p)}$. By integrating the probabilities over
the impact parameter $b$, the production cross section for a fragment
composed of $N$ neutrons and $Z$ protons is evaluated as
\begin{eqnarray}
\sigma (N,Z) = 2\pi \int_0^\infty b\; P_{N,Z}(b)\, db.
\label{sigmatot}
\end{eqnarray}
We perform TDHF calculations of the $^{238}$U+$^{124}$Sn
collision for impact parameters from 0 to 10~fm. For an impact
parameter region from 0 to 5~fm and from 5 to 10~fm, we achieve
calculations in 0.5-fm and 1-fm steps, respectively. We evaluate
the cross section by numerical quadrature according to Eq.~(\ref{sigmatot}).

The ground state of $^{238}$U is prolately deformed with
$\beta \sim 0.27$ and the ground state of $^{124}$Sn is
oblately deformed with $\beta \sim 0.11$. We performed TDHF
calculations for three initial configurations characterized by different
orientations of $^{238}$U: The symmetry axis of $^{238}$U set
parallel to the $x$-axis (parallel to the collision axis), $y$-axis (parallel
to the impact parameter vector), and $z$-axis (perpendicular to the
collision plane). The symmetry axis of a slightly deformed $^{124}$Sn
is always set parallel to the $z$-axis. For a quantitative comparison
with the measured cross sections, we should average over all possible
orientations. However, since the orientation average requires excessive
computational costs, we have not yet performed it. Below, we show
cross sections for each of the three initial conditions without the
average, Eq.~(\ref{sigmatot}).

\begin{figure}[t]
   \begin{center}
   \includegraphics[width=15.4cm]{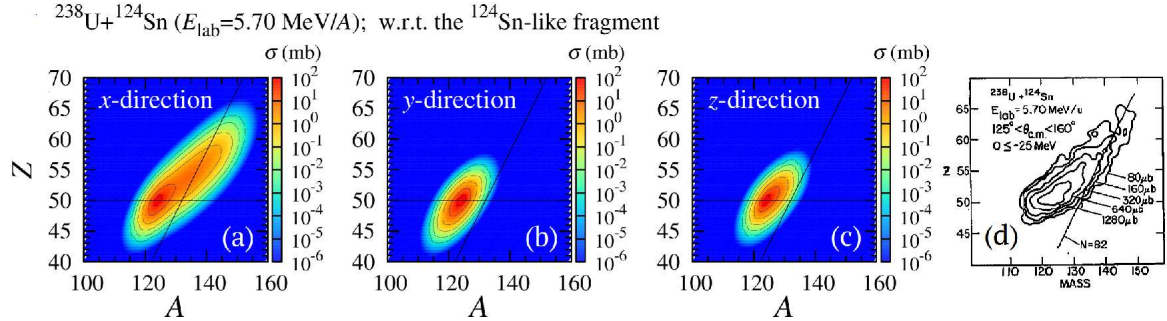}
   \end{center}\vspace{-1.5mm}
   \caption{
   Production cross sections of $^{124}$Sn-like fragments in
   $^{238}$U+$^{124}$Sn collisions at $E_\mathrm{lab}=5.7$~MeV/$A$
   are shown in the $A$-$Z$ plane. (a-c): Results of the TDHF calculations for
   three different relative orientations. (d): Experimentally measured
   cross sections (Figure has been taken from Ref.~\cite{Mayer238U}).
   }\vspace{-4.5mm}
   \label{NTCS}
\end{figure}

In Fig.~\ref{NTCS}, we show production cross sections of $^{124}$Sn-like
fragments in the $A$-$Z$ plane. In Fig.~\ref{NTCS}~(a), (b), and (c),
we show cross sections calculated by Eq.~(\ref{sigmatot}) for different
initial configurations. From the results shown in the panels (a), (b), and (c),
we find that the distributions of the calculated cross sections depend much
on the initial orientations of the deformed $^{238}$U. 

\begin{wrapfigure}{r}{6.5cm}
\vspace{-4.5mm}
   \begin{center}
   \includegraphics[width=6.5cm]{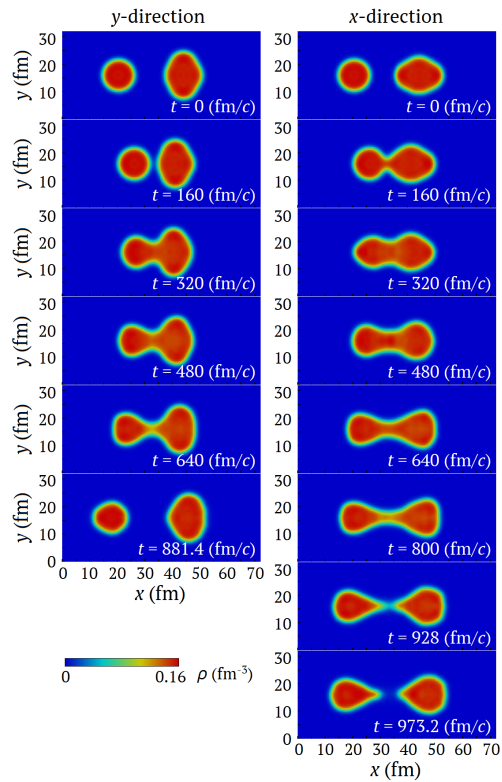}
   \end{center}\vspace{-1mm}
   \caption{
   Time evolutions of the density distribution in the collision plane for head-on collisions
   of $^{238}$U+$^{124}$Sn at different initial configurations. (Left panels):
   A case of the side collision in which the symmetry axis of $^{238}$U
   is set parallel to the $y$-axis. (Right panels): A case of the tip collision in which
   the symmetry axis of $^{238}$U is set parallel to the $x$-axis.
   }\vspace{-5mm}
   \label{rho}
\end{wrapfigure}

When the symmetry axis of $^{238}$U is set parallel to the collision
axis ($x$-direction in panel (a)), we find abundant cross sections widely
spreading in the $A$-$Z$ plane. For a fragment $_{\;\:44}^{116}$Ru$_{72}^{}$
produced by a transfer of two neutrons and six protons from 
$_{\;\:50}^{124}$Sn$_{74}^{}$, we find a cross section of $10^{-3}$~mb. 
For a fragment $_{\;\:64}^{150}$Gd$_{86}^{}$ produced by a transfer
of twelve neutrons and fourteen protons to $_{\;\:50}^{124}$Sn$_{74}^{}$,
the cross section is again the same order of magnitude, $10^{-3}$~mb.

When the symmetry axis of $^{238}$U is set perpendicular to the
collision axis (symmetry axis in $y$- and $z$-directions, shown in
panels (b) and (c), respectively), the calculated cross sections do
not so much extend in the $A$-$Z$ plane compared to the case of
$x$-direction shown in panel (a). Cross sections producing lighter nuclei
in the transfer from $^{124}$Sn to $^{238}$U are almost the same
as those in the $x$-direction case. However, cross sections to produce
heavier nuclei in the transfer from $^{238}$U to $^{124}$Sn is
substantially suppressed compared to the $x$-direction case. For
example, we find a cross section of $10^{-3}$~mb for the production
of $_{\;\:56}^{134}$Ba$_{78}^{}$ which corresponds to a transfer
of four neutrons and six protons to $_{\;\:50}^{124}$Sn$_{74}^{}$.
The number of transferred nucleons with the similar magnitude of cross
section is  much smaller than the cross section shown in panel $(a)$.

To obtain intuitive pictures for the reaction dynamics, we show in
Fig.~\ref{rho} time evolutions of the calculated density distribution
in the collision plane ($xy$-plane). We show results of head-on collisions
($b=0$~fm) with two different initial orientations. In the $x$-direction
case, the symmetry axis of $^{238}$U is set parallel to the collision axis.
In the $y$-direction case, symmetry axis of $^{238}$U is set perpendicular
to the collision axis. The top panels show initial configurations. 
We show several snapshots below.

In both $x$- and $y$-direction cases, two nuclei touch at around
320~fm/$c$. In the $x$-direction case (right panels), a thick neck is
developed between the two colliding nuclei forming an elongated
di-nuclear system (480-800 fm/$c$). When the di-nuclear system
dissociates ($\sim 928$~fm/$c$), the neck is cut at a position closer
to the larger fragment. Consequently, a lot of nucleons in the neck
region are absorbed by the smaller fragment. Since the neck region is
composed of both neutrons and protons, the absorption of nucleons
in the neck region results in the transfer of both neutrons and protons
in the same direction. We find that about eleven neutrons and seven
protons are transferred in average in this reaction, producing fragments
resembling $_{\;\:57}^{142}$La$_{85}^{}$ and 
$_{\;\:85}^{219}$At$_{134}^{}$. In the $y$-direction case
(left panels), on the other hand, the neck is not so much developed
compared to the $x$-direction case (320-640~fm/$c$). As a result,
only one-neutron and one-proton are transferred on average from
$^{238}$U to $^{124}$Sn.

In Fig.~\ref{NTCS}~(d), we show measured production cross sections
for $^{124}$Sn-like fragments in the $^{238}$U+$^{124}$Sn collisions
reported in Ref.~\cite{Mayer238U}. As seen in Fig.~\ref{NTCS}~(d),
measured cross sections extend to the mass number $A \sim 148$
and the proton number $Z \sim 64$. It corresponds to a transfer
of eight neutrons and fourteen protons from $^{238}$U to $^{124}$Sn.
As seen in Fig.~\ref{NTCS}~(a), (b), and (c), the large number of
transferred nucleons from $^{238}$U to $^{124}$Sn in the
measurement can only be explained by the $x$-direction configuration,
the tip collision of a deformed $^{238}$U, among the examined
three configurations. Our TDHF calculations strongly suggest that the
large number of transferred nucleons, more than ten protons, from
$^{238}$U to $^{124}$Sn in the measured MNT processes can only
be explained in the tip-collision-induced transfer, associated with the
formation and absorption of the elongated thick neck during the collision.

\section{Concluding Remarks}\vspace{-1mm}

We have investigated MNT processes in $^{238}$U+$^{124}$Sn
collisions at around the Coulomb barrier employing the TDHF theory
with the particle-number projection method. For this reaction,
substantial cross sections of MNT processes accompanying more than
ten protons were measured experimentally. Since $^{238}$U is a
prolately deformed nucleus, we performed reaction calculations for
three initial orientations of the deformed $^{238}$U. The symmetry
axis of $^{124}$Sn, which is slightly deformed in oblate shape, is
always set perpendicular to the collision plane. From the calculation,
we have found that the MNT processes accompanying more than
ten protons can only be explained in the tip collision in which the
symmetry axis of $^{238}$U is parallel to the incident direction.
An extended neck is formed in the tip collision, and the breaking
of the neck is responsible for the MNT accompanying a large number
of protons. To clarify reaction mechanisms further, more detailed
investigations such as projectile-target combination and/or incident
energy dependence as well as effects of particle evaporation should
be considered. A study along this line is now in progress and results
will be discussed in a forthcoming paper.

\section*{Acknowledgments}\vspace{-2mm}
This research used computational resources of the HPCI system
provided by Information Initiative Center, Hokkaido University,
through the HPCI System Research Project (Project ID: hp140010).
A part of calculations was carried out using computational resources
of the COMA (PACS-IX) system at the Center for Computational
Sciences at the University of Tsukuba. This work was supported
by the Japan Society of the Promotion of Science (JSPS)
Grants-in-Aid for Scientific Research Grant Numbers
23340113 and 25104702, and by the JSPS Grant-in-Aid for
JSPS Fellows Grant Number 25-241.

\end{document}